# The Survey of Data Mining Applications And Feature Scope

Neelamadhab Padhy[1], Dr. Pragnyaban Mishra [2], and Rasmita Panigrahi[3]

[1] Asst. Professor, Gandhi Institute of Engineering and Technology, GIET, Gunupur
`neela.mbamtech@gmail.com`
Dept. of Computer Science, CMJ University, Meghalaya (Silong) a research Scholar
[2] Associate Professor, Gandhi Institute of Engineering and Technology, GIET, Gunupur
`pragnyaban@gmail.com`
[3] Lecturer, Gandhi Institute of Engineering and Technology, GIET, Gunupur
`rasmi.mcamtech@gmail.com`

*Abstract*

*In this paper we have focused a variety of techniques, approaches and different areas of the research which are helpful and marked as the important field of data mining Technologies. As we are aware that many **MNC's** and large organizations are operated in different places of the different countries. Each place of operation may generate large volumes of data. Corporate decision makers require access from all such sources and take strategic decisions .The data warehouse is used in the significant business value by improving the effectiveness of managerial decision-making. In an uncertain and highly competitive business environment, the value of strategic information systems such as these are easily recognized however in today's business environment, efficiency or speed is not the only key for competitiveness. This type of huge amount of data's are available in the form of tera- to peta-bytes which has drastically changed in the areas of science and engineering. To analyze, manage and make a decision of such type of huge amount of data we need techniques called the data mining which will transforming in many fields. This paper imparts more number of applications of the data mining and also o focuses scope of the data mining which will helpful in the further research.*

*Keywords*

*Data mining task, Data mining life cycle , Visualization of the data mining model , Data mining Methods, Data mining applications,*

## 1. INTRODUCTION

In the 21$^{st}$ century the human beings are used in the different technologies to adequate in the society . Each and every day the human beings are using the vast data and these data are in the different fields .It may be in the form of documents, may be graphical formats ,may be the video ,may be records (varying array ) .As the data are available in the different formats so that the proper action to be taken. Not only to analyze these data but also take a good decision and maintain the data .As and when the customer will required the data should be retrieved from the database and make the better decision .This technique is actually we called as a data mining or Knowledge Hub or simply KDD(Knowledge Discovery Process).The important reason that attracted a great deal of attention in information technology the discovery of useful information from large collections of data industry towards field of "Data mining" is due to

DOI : 10.5121/ijcseit.2012.2303 43

the perception of *"we are data rich but information poor"*. There is huge volume of data but we hardly able to turn them in to useful information and knowledge for managerial decision making in business. To generate information it requires massive collection of data. It may be different formats like audio/video, numbers, text, figures, Hypertext formats . To take complete advantage of data; the data retrieval is simply not enough, it requires a tool for automatic summarization of data, extraction of the essence of information stored, and the discovery of patterns in raw data. With the enormous amount of data stored in files, databases, and other repositories, it is increasingly important, to develop powerful tool for analysis and interpretation of such data and for the extraction of interesting knowledge that could help in decision-making. The only answer to all above is 'Data Mining'. Data mining is the extraction of hidden predictive information from large databases; it is a powerful technology with great potential to help organizations focus on the most important information in their data warehouses [1,2,3,4]. Data mining tools predict future trends and behaviors, helps organizations to make proactive knowledge-driven decisions [2]. The automated, prospective analyses offered by data mining move beyond the analyses of past events provided by prospective   tools typical of decision support systems. Data mining tools can answer the questions that traditionally were too time consuming to resolve. They prepare databases for finding hidden patterns, finding predictive information that experts may miss because it lies outside their expectations.

Data mining, popularly known as Knowledge Discovery in Databases (KDD), it is the nontrivial extraction of implicit, previously unknown and potentially useful information from data in databases [3, 5]**.** It is actually the process of finding the hidden information/pattern of the repositories **.[1,3,5].**

This paper describes 7 sections .Section 1 is completely introduction where you will get huge information about the data mining concept. Section 2 describes the data mining task which describes that how the data will be store, how to retrieve, how to analyze the data. .Section 3 focuses the data mining classification tasks .section 4 provides the data mining life cycles. Section 5 describes visualization of the data model and it involves extracting the hidden information as we as **we have proposed the new way to define KDD Process**. Section 6 describes shortly, some of the popular data mining methods. The chapter 7 is the heart of the paper, we have reviewed applications and we propose feature directions some of data mining applications. We have added the scope of the data mining applications so that the researcher can pin pointed the following areas.

## 2. The Data Mining Task

The data mining tasks are of d*i*fferent types depending on the use of data mining result the data mining tasks are classified as[1,2]:

### 2.1 Exploratory Data Analysis:

In the repositories vast amount of information's are available .This data mining task will serve the two purposes
(i).With out the knowledge for what the customer is searching, then
(ii) It analyze the data
These techniques are interactive and visual to the customer.

### 2.2  Descriptive Modeling:

It describe all the data, it includes models for overall probability distribution of the data, partitioning of the   p-dimensional space into groups and models describing the relationships between the variables.

44

**2.3 Predictive Modeling:**

This model permits the value of one variable to be predicted from the known values of other variables.

**2.4. Discovering Patterns and Rules:**

This task is primarily used to find the hidden pattern as well as to discover the pattern in the cluster. In a cluster a number of patterns of different size and clusters are available .The aim of this task is "how best we will detect the patterns" .This can be accomplished by using rule induction and many more techniques in the data mining algorithm like(K-Means /K-Medoids) .These are called the clustering algorithm.

**2.5 Retrieval by Content:**

The primary objective of this task is to find the data sets of frequently used in the for audio/video as well as images It is finding pattern similar to the pattern of interest in the data set

## 3. Types of Data Mining System:

Data mining systems can be categorized according to various criteria the classification is as follows[3]:

**3.1 Classification of data mining systems according to the type of data source mined:**

In an organization a huge amount of data's are available where we need to classify these data but these are available most of times in a similar fashion. we need to classify these data according to its type(maybe audio/video ,text format etc)

**3.2 Classification of data mining systems according to the data model:**
There are so many number of data mining models ( Relational data model, Object Model, Object Oriented data Model, Hierarchical data Model/W data model )are available and each and every model we are using the different data .According to these data model the data mining system classify the data in the model.

**3.3 Classification of data mining systems according to the kind of knowledge discovered:**

This classification based on the kind of knowledge discovered or data mining functionalities, such as characterization, discrimination, association, classification, clustering, etc. Some systems tend to be comprehensive systems offering several data mining functionalities together.

**3.4 Classification of data mining systems according to mining techniques used:**

This classification is according to the data analysis approach used such as machine learning, neural networks, genetic algorithms, statistics, visualization, database oriented or data warehouse-oriented, etc.

The classification can also take into account the degree of user interaction involved in the data mining process such as query-driven systems, interactive exploratory systems, or autonomous systems. A comprehensive system would provide a wide variety of data mining techniques to fit different situations and options, and offer different degrees of user interaction.

## 4. Data Mining Life Cycle:

The life cycle of a data mining project consists of six phases[2,4]. The sequence of the phases is not rigid. Moving back and forth between different phases is always required. It depends on the outcome of each phase. The main phases are:

### 4.1. Business Understanding:

This phase focuses on understanding the project objectives and requirements from a business perspective, then converting this knowledge into a data mining problem definition and a preliminary plan designed to achieve the objectives.

### 4.2 Data Understanding:

It starts with an initial data collection, to get familiar with the data, to identify data quality problems, to discover first insights into the data or to detect interesting subsets to form hypotheses for hidden information.

### 4.3 Data Preparation:

In this stage , it collects all the different data sets and construct the varieties of the activities basing on the initial raw data

### 4.4 Modeling:

In this phase, various modeling techniques are selected and applied and their parameters are calibrated to optimal values.

### 4.5 Evaluation:

In this stage the model is thoroughly evaluated and reviewed. The steps executed to construct the model to be certain it properly achieves the business objectives. At the end of this phase, a decision on the use of the data mining results should be reached.

### 4.6 Deployment:

The purpose of the model is to increase knowledge of the data, the knowledge gained will need to be organized and presented in a way that the customer can use it. The deployment phase can be as simple as generating a report or as complex as implementing a repeatable data mining process across the enterprise.

## 5. Visualizing Data Mining Model

The main objective of data visualization is the overall idea about the data mining model .In data mining most of the times we are retrieving the data from the repositories which are in the hidden form. This is the difficult task for a user. So this visualization of the data mining model helps us to provide  utmost levels of understanding and trust. Because the user does not know beforehand what the data mining process has discovered, it is a much bigger leap to take the output of the system and translate it into an actionable solution to a business problem. The data mining models are of two types [1,2,6,45]: Predictive and Descriptive.

The predictive model makes prediction about unknown data values by using the known values. Ex. Classification, Regression, Time series analysis, Prediction etc. The descriptive model identifies the patterns or relationships in data and explores the properties of the data examined. Ex. Clustering, Summarization, Association rule, Sequence discovery etc.

Many of the data mining applications are aimed to predict the future state of the data. Prediction is the process of analyzing the current and past states of the attribute and prediction of its future state. Classification is a technique of mapping the target data to the predefined groups or classes, this is a supervise learning because the classes are predefined before the examination of the target data. The regression involves the learning of function that map data item to real valued prediction variable. In the time series analysis the value of an attribute is examined as it varies over time. In time series analysis is used for many statistical techniques which will analyze the time-series data such as auto regression methods etc.It is some times used in the two type of modeling (1) ARIMA (II)Long-memory time-series modeling .

The term clustering means analyzes the different data objects without consulting a known class levels. It is also referred to as unsupervised learning or segmentation. It is the partitioning or segmentation of the data in to groups or clusters. The clusters are defined by studying the behavior of the data by the domain experts. The term segmentation is used in very specific context; it is a process of partitioning of database into disjoint grouping of similar tuples. Summarization is the technique of presenting the summarize information from the data. The association rule finds the association between the different attributes. Association rule mining is a two-step process: Finding all frequent item sets, Generating strong association rules from the frequent item sets. Sequence discovery is a process of finding the sequence patterns in data. This sequence can be used to understand the trend.

**New way to define the KDD Process:**

We have found the broader meaning of the followings Data, patterns, Process, Valid, Novel, and Useful Understandable. Of KDD. The Knowledge discovery in databases is the non-trivial process of identifying valid, novel, potentially useful, and ultimately understandable patterns in data.

**Table 1.1 to describe the new form the word**

| Data | A set of facts, *F*. |
| --- | --- |
| Pattern | An expression *E* in a language *L* describing facts in a subset $F_E$ of *F*. |
| Process | It means different operations associated with the KDD .The operations involving preparation of the data ,searching the different patterns , Judging the knowledge and evaluation etc. |
| Valid | Those patterns which are discovered that are completely new one and which can be used feature |
| Novel | Derive the hidden patterns |
| Useful | Newly discovered patterns should be used for different actions . |

## 6. Data Mining Methods:

**Some of the popular data mining methods are as follows:**

6.1　　Decision Trees and Rules
6.2　　Nonlinear Regression and Classification Methods
6.3　　Example-based Methods
6.4　　Probabilistic Graphical Dependency Models
6.5　　Relational Learning Models

We found these are some famous data mining methods are broadly classified as: On-Line Analytical Processing ,(OLAP), Classification, Clustering, Association Rule Mining, Temporal Data Mining, Time Series Analysis, Spatial Mining, Web Mining etc. These methods use different types of algorithms and data. The data source can be data warehouse, database, flat file or text file. The algorithms may be Statistical Algorithms, Decision Tree based, Nearest Neighbor, Neural Network based, Genetic Algorithms based, Ruled based, Support Vector Machine etc. Generally the data mining algorithms are fully dependent of the two factors these are

(i)　　　which type of data sets are using
(ii)　　　what type of requirements of the user

Basing upon the above two factors the data mining algorithms are used.A knowledge discovery (KD) process involves preprocessing data, choosing a data-mining algorithm, and post processing the mining results. The Intelligent Discovery Assistants [7] (IDA), helps users in applying valid knowledge discovery processes. The IDA can provide users with three benefits:

- A systematic enumeration of valid knowledge discovery processes;
- Effective rankings of valid processes by different criteria, which help to choose between the options;
- An infrastructure for sharing knowledge, which leads to network externalities.

Several other attempts have been made to automate this process and design of a generalized data mining tool that posse's intelligence to select the data and data mining algorithms and up to some extent the knowledge discovery.

## 7. Data Mining Applications:

In this section, we have focused some of the applications of data mining and its techniques are analyzed respectively Order.

### 7.1 Data Mining Applications in Healthcare

Data mining applications in health can have tremendous potential and usefulness [60]. However, the success of healthcare data mining hinges on the availability of clean healthcare data. In this respect, it is critical that the healthcare industry look into how data can be better captured, stored, prepared and mined. Possible directions include the standardization of clinical vocabulary and the sharing of data across organizations to enhance the benefits of healthcare data mining applications

### 7.1.1 Future Directions of Health care system through Data Mining Tools

As healthcare data are not limited to just quantitative data (e.g., doctor's notes or clinical records), it is necessary to also explore the use of text mining to expand the scope and nature of what healthcare data mining can currently do. This is specially used to mixed all the data and then mining the text. It is also useful to look into how images (e.g., MRI scans) can be brought into healthcare data mining applications. It is noted that progress has been made in these areas

### 7.2 Data mining is used for market basket analysis

Data mining technique is used in MBA(Market Basket Analysis).When the customer want to buying some products then this technique helps us finding the associations between different items that the customer put in their shopping buckets. Here the discovery of such associations that promotes the business technique .In this way the retailers uses the data mining technique so that they can identify that which customers intension (buying the different pattern).In this way this technique is used for profits of the business and also helps to purchase the related items.

### 7.3 The data mining is used an emerging trends in the education system [57, 58] in the whole world

In Indian culture most of the parents are uneducated .The main aim of in Indian government is the quality education not for quantity. But the day by day the education systems are changed and in the $21^{st}$ century a huge number of universalities are established by the order of UGC. As the numbers of universities are established side by side, each and every day a millennium of students are enrolls across the country. With huge number of higher education aspirants, we believe that data mining technology can help bridging knowledge gap in higher educational systems. The hidden patterns, associations, and anomalies that are discovered by data mining techniques from educational data can improve decision making processes in higher educational systems. This improvement can bring advantages such as maximizing educational system efficiency, decreasing student's drop-out rate, and increasing student's promotion rate, increasing student's retention rate in, increasing student's transition rate, increasing educational improvement ratio, increasing student's success, increasing student's learning outcome, and reducing the cost of system processes. In this current era we are using the KDD and the data mining tools for extracting the knowledge this knowledge can be used for improving the quality of education .The decisions tree classification is used in this type of applications.

### 7.4 Data mining is now used in many different areas in manufacturing engineering [59]

When we retrieve the data from manufacturing system then the customer is to use these data for different purposes like to find the errors in the data ,to enhance the design methodology ,to make the good quality of the data ,how best the data can be supported for making the decision . But most of time the data can be first analyzed then after find the hidden patterns which will be control the manufacturing process which will further enhance the quality of the products .Since the importance of data mining in manufacturing has clearly increased over the last 20 years, it is now appropriate to critically review its history and Application

#### 7.4.1 Future Directions in the manufacturing Engineering through the Data mining Tools

It is very tedious task to mine the manufacturing data .Generally when we mine the data in the manufacturing, we dose not give more important to the quality of the rules .After mining those knowledge which has generated is very difficult because relationship identification is too complex to understand. That's why we need the further to enhance the research methodology to know the proper knowledge. The new methodology was proposed i.e CRISP-DM which will provides the high level detail steps of instructions for using the data mining in the engineering field. Further research is needed to develop generic guidelines for a variety of different data and types of problems, which are commonly faced by manufacturing engineering industry
.

### 7.5 Data Mining Applications can be generic or domain specific.

Data mining system can be applied for generic or domain specific . Some generic data mining applications cannot take its own these decisions but guide users for selection of data, selection of data mining method and for the interpretation of the results. The multi agent based data mining application [8, 10] has capability of automatic selection of data mining technique to be applied. The Multi Agent System used at different levels [8]: First, at the level of concept hierarchy definition then at the result level to present the best adapted decision to the user. This decision is stored in knowledge Base to use in a later decision-making. Multi Agent System Tool used for generic data mining system development [10] uses different agents to perform different tasks.

### 7.6 A multi-tier data mining system is proposed to enhance the performance of the data mining process [9].

It has basic components like user interface, data mining services, data access services and the data. There are three different architectures presented for the data mining system namely one-tire, Two-tire and Three-tire architecture. Generic system required to integrate as many learning algorithms as possible and decides the most appropriate algorithm to use. CORBA (Common Object Request Broker Architecture) has features like: Integration of different applications coded in any programming language considerably easy. It allows reusability in a feasible way and finally it makes possible to build large and scalable system. The data mining system architecture based on CORBA is given by Object Management Group [10] has all characteristics to accomplish a distributed and object oriented computation. A data-centric focus and automated methodologies makes data mining accessible to no experts [11]. The use of high-level interfaces can implement the automated methodologies that hide the data mining concepts away from the users. A data-centric design hides away all the details of mining methodology and exposes them through high-level tasks that are goal-oriented. These goal-oriented tasks are implemented using data-centric APIs. This design makes data mining task like other types of queries that users perform on the data. In data mining better results could be obtained if large data is available. It leads to the merging and linking of local databases. A new data-mining architecture based on Internet technology addressed this problem. [12] The context factor plays vital role in the success of data mining. The importance and meaning of same data in the different context is different. A data in one context is very important may not be much important in other context. A context-aware data-mining framework filters useful and interesting context factors, and can produce accurate and precise prediction using those factors[46].

### 7.7 Application of Data Mining techniques in CRM

Data mining technique is used in CRM .Now a days it is one of the hot topic to research in the industry because CRM have attracted both the practitioners and academics. It aims to give a research summary on the application of data mining in the CRM domain and techniques which are most often used. Although this review cannot claim to be exhaustive, it does provide reasonable insights and shows the incidence of research on this subject. The results presented in this paper have several important implications: Research on the application of data mining in CRM will increase significantly in the future based on past publication rates and the increasing interest in the area. The majority of the reviewed articles relate to customer retention [49]

### 7.8 The Domain Specific Applications

The domain specific applications are focused to use the domain specific data and data mining algorithm that targeted for specific objective. The applications studied in this context are aimed to generate the specific knowledge. In the different domains the data generating sources generate different type of data. Data can be from a simple text, numbers to more complex audio-video data. To mine the patterns and thus knowledge from this data, different types of data mining algorithms are used. The collection and selection of context specific data and applying the data mining algorithm to generate the context specific knowledge is thus a skillful job. In many domains specific data mining applications the domain experts plays vital role to mine useful knowledge.

In the identification of foreign-accented French the audio files were used and the best 20 data mining algorithms were applied[13] the Logistic Regression model found the most robust algorithm than other algorithm.

**7.9 In language research and language** engineering much time extra linguistic information is needed about a text. A linguistic profile that contains large number of linguistic features can be generated from text file automatically using data mining [14]. This technique found quite effective for authorship verification and recognition. A profiling system using combination of lexical and syntactic features shows 97% accuracy in selecting correct author for the text. The linguistic profiling of text effectively used to control the quality of language and for the automatic language verification.[15] This method verifies automatically the text is of native quality. The results show that language verification is indeed possible.

### 7.10 In Medical Science

In medical science there is large scope for application of data mining. Diagnosis of diesis, health care, patient profiling and history generation etc. are the few examples. Mammography is the method used in breast cancer detection. Radiologists face lot of difficulties in detection of tumors that's why CAM(Computer Aided Methods) could helps to the medical staff . So that they can produce the good quality of the result detection [16]. The neural networks with back-propagation and association rule mining used for tumor classification in mammograms. The data mining effectively used in the diagnosis of lung abnormality that may be cancerous or benign [17]. The data mining algorithms significantly reduce patient's risks and diagnosis costs. Using the prediction algorithms the observed prediction accuracy was 100% for 91.3% cases. The use of data mining in health care is the widely used application of data mining. The medical data is complex and difficult to analyze. A REMIND (Reliable Extraction and Meaningful Inference from Non-structured Data) system [21] integrates the structured and unstructured clinical data in patient records to automatically create high quality structured clinical data. To adopt the high quality technique, we can mined the existing patient records to support guidelines and give compliance to improve patient care. [21]

### 7.11 Data Mining methods are used in the Web Education

Data mining methods are used in the web Education which is used to improve courseware. The relationships are discovered among the usage data picked up during students' sessions. This knowledge is very useful for the teacher or the author of the course, who could decide what modifications will be the most appropriate to improve the effectiveness of the course. [42].In the 21$^{st}$ century the beginners are using the data mining techniques which is one of the best learning method in this era[41]. This makes it possible to increase the awareness of learners. Web Education which will rapidly growth in the application of data mining methods to educational chats which is both feasible and can be improvement in learning environments in the 21$^{st}$ century.

## 7.12 Credit Scoring

Credit scoring has become very important issue due to the recent growth of the credit industry, so the credit department of the bank faces the huge numbers of consumers' credit data to process, but it is impossible analyzing this huge amount of data both in economic and manpower terms. In this study we reviewed the papers which have applied data mining methods in credit risk evaluation problem. Ten data mining technique which were most used method in the credit risk evaluation context were extracted, and then we searched almost all papers which had focused on these ten methods form 2000 to 2011. It is concluded that the support vector machine has been widely applied in recent years and which is one of the best technique.. Since to improve the performance of this model, it is necessary a method for reduction the feature subset, many hybrid SVM based model are proposed. Moreover the hybrid models have been attended in the last decade because of its enjoying from advantages of two or more models. Many of these proposed models can only classify customers into two classes "good" or "bad" ones. Several single and hybrid data mining methods are applied for credit scoring problem [50], [51], [53], [54], [55].The most used applied methods for doing credit scoring task are derived from classification technique. Generally classification is used when we predict some thing which is possible by using the previous available information. It is one type of methods which can be defined as classification where the members of a given set of instances into some groups where the different types of characteristics are to be made. Classification task is very suited to data mining methods and techniques

## 7.13 The Intrusion Detection in the Network

The intrusion detection in the Network is very difficult and needs a very close watch on the data traffic. The intrusion detection plays an essential role in computer security. The classification method of data mining is used to classify the network traffic normal traffic or abnormal traffic.[26].  If any TCP header does not belong to any of the existing TCP header clusters, then it can be considered as anomaly.

## 7.14 A malicious Executable is Threat

A malicious executable is threat to system's security, it damage a system or obtaining sensitive information without the user's permission. The data mining methods used to accurately detect malicious executables before they run[25]. Classification algorithms RIPPER, Naive Bayes, and a Multi-Classifier system are used to detect new malicious executables. This classifier had shown detection rate 97.76%.

## 7.15 Sports data Mining :

The data mining and its technique is used for an application of Sports center. Data mining is not only use in the business purposes but also it used in the sports .In the world, a huge number of games are available where each and every day the national and international games are to be scheduled, where a huge number of data's are to be maintained .The data mining tools are applied to give the information as and when we required. The open source data mining tools like WEKA and RAPID MINER frequently used for sport. This means that users can run their data through one of the built-in algorithms, see what results come out, and then run it through a different algorithm to see if anything different stands out. As these programs' are available in the form of open source in  nature, that's why the users are frequently  to modify the source code, so that other can get the updated information [56 ] . In the sports world the vast amounts of statistics are collected for each player, team, game, and season. In the game sports the data's are available in the form of statistical form where data mining can be used and discover the patterns, these patterns are often  used to predict the future forecast . Data   mining can be used for scouting, prediction of performance, selection of players, coaching and training and for the strategy planning [34]. The data mining techniques are used to determine the best or the most optimal squad to represent a team in a team sport in a season, tour or game.[44] The 'Cy Young Award'[30] has been presented annually to the best pitcher in the major league of baseball. The award is based largely on statistics compiled over the course of the baseball season. A Bayesian classifier is developed to predict Cy Young Award winners in American major league baseball.

### 7.16 The Intelligence Agencies

The Intelligence Agencies collect and analyze information to investigate terrorist activities. One challenge to law enforcement and intelligent agencies is the difficulty of analyzing large volume of data involve in criminal and terrorist activities. Now a days the intelligence agency are using the sophisticated data mining algorithms which makes it easy, to handle the very large data bases databases for organizations. The different data mining techniques are used in crime data mining. [ 33],[37] .Though the organization's have used large data bases but data mining helps us to generate the different types of information in the organization like personal details of the persons along with, vehicle details .In data mining the Clustering techniques is used (Association rule mining) for the different objects(like persons, organizations, vehicles etc.) in crime records. Not only data mining detects but also analyzes the crime data. The classification technique is also used to detect email spamming and also find person who has given the mail. String comparator is used to detect deceptive information in criminal record.

### 7.17 The data mining system implemented at the Internal Revenue Service

The data mining system implemented at the Internal Revenue Service to identify high-income individuals engaged in abusive tax shelters [23] show significantly good results. The major lines of investigation included visualization of the relationships and data mining to identify and rank possibly abusive tax avoidance transactions. To enhance the quality of product data mining techniques can be used effectively. The data mining technology SAS/EM is used to discover the rules those are unknown before and it can improve the quality of products and decrease the cost. A regression model and the neural network model when applied for this purpose given accuracy above 80%. [31] The neural network model found better than the regression model.

### 7.18 E-commerce is also the most prospective

E-commerce is also the most prospective domain for data mining [39]. It is ideal because many of the ingredients required for successful data mining are easily available: data records are plentiful, electronic collection provides reliable data, insight can easily be turned into action, and return on investment can be measured. The integration of e-commerce and data mining significantly improve the results and guide the users in generating knowledge and making correct business decisions. This integration effectively solves several major problems associated with horizontal data mining tools including the enormous effort required in pre-processing of the data before it can be used for mining, and making the results of mining actionable.

### 7.19 The Digital Library Retrieves

The data mining application can be used in the field of the Digital Library where the user will finds or collects, stores and preserves the data which are in the form of digital mode. The advent of electronic resources and their increased use in libraries has brought about significant changes in Library [40]. The data and information are available in the different formats. These formats include Text, Images, Video, Audio, Picture, Maps, etc. therefore digital library is a suitable domain for application of data mining.

### 7.20 The prediction in engineering applications

The prediction in engineering applications was treated effectively by a data mining approach[17]. The prediction problems like the cost estimation problem in engineering, the problem of engineering design that involves decisions where parameters, actions, components, and so on are selected. data mining technique is used for the variety of the parameters in the field of engineering applications like prior data .Once we gather the data then we can generate the different models ,algorithms which will predict to different characteristic. The data mining algorithm applied on the test file with nine features has produced 100% correct predictions. Several other applications studied in this context.

## 8. The Scope of Data Mining

Data mining derives its name from the similarities between searching for valuable business information in a large database for example, finding linked products in gigabytes of store scanner data and mining a mountain for a vein of valuable ore. Both processes require either sifting through an immense amount of material, or intelligently probing it to find exactly where the value resides. Given databases of sufficient size and quality, data mining technology can generate new business opportunities by providing these capabilities:

### 8.1 Automated prediction of trends and behaviors.

Data mining automates the process of finding predictive information in large databases. Questions that traditionally required extensive hands-on analysis can now be answered directly from the data quickly. A typical example of a predictive problem is targeted marketing. Data mining uses data on past promotional mailings to identify the targets most likely to maximize return on investment in future mailings. Other predictive problems include forecasting bankruptcy and other forms of default, and identifying segments of a population likely to respond similarly to given events.

- **Artificial neural networks**:

Non-linear predictive models that learn through training and resemble biological neural networks in structure.

- **Decision trees**:

Tree-shaped structures that represent sets of decisions. These decisions generate rules for the classification of a dataset. Specific decision tree methods include Classification and Regression Trees (CART) and Chi Square Automatic Interaction Detection (CHAID) .

- **Genetic algorithms**:

Optimization techniques that use process such as genetic combination, mutation, and natural selection in a design based on the concepts of evolution.

- **Nearest neighbor method**:

A technique that classifies each record in a dataset based on a combination of the classes of the k record(s) most similar to it in a historical dataset (where k ³ 1). Sometimes called the k-nearest neighbor technique.

- **Rule induction**:

The extraction of useful if-then rules from data based on statistical significance. Many of these technologies have been in use for more than a decade in specialized analysis tools that work with relatively small volumes of data. These capabilities are now evolving to integrate directly with industry-standard data warehouse and OLAP platforms.

## 9. Conclusion:

In this paper we briefly reviewed the various data mining applications. This review would be helpful to researchers to focus on the various issues of data mining. In future course, we will review the various classification algorithms and significance of evolutionary computing (genetic programming) approach in designing of efficient classification algorithms for data mining. Most of the previous studies on data mining applications in various fields use the variety of data types range from text to images and stores in variety of databases and data structures. The different methods of data mining are used to extract the patterns and thus the knowledge from this variety databases. Selection of data and methods for data mining is an important task in this process and needs the knowledge of the domain. Several attempts have been made to design and develop the generic data mining system but no system found completely generic. Thus, for every domain the domain expert's assistant is mandatory. The domain experts shall be guided by the system to effectively apply their knowledge for the use of data mining systems to generate required knowledge. The domain experts are required to determine the variety of data that should be collected in the specific problem domain, selection of specific data for data mining, cleaning and transformation of data, extracting patterns for knowledge generation and finally interpretation of the patterns and knowledge generation. Most of the domain specific data mining applications show accuracy above 90%. The generic data mining applications are having the limitations. From the study of various data mining applications it is observed that, no application called generic application is 100 % generic. The intelligent interfaces and intelligent agents up to some extent make the application generic but have limitations. The domain experts play important role in the different stages of data mining. The decisions at different stages are influenced by the factors like domain and data details, aim of the data mining, and the context parameters. The domain specific applications are aimed to extract specific knowledge. The domain experts by considering the user's requirements and other context parameters guide the system. The results yield from the domain specific applications are more accurate and useful. Therefore it is conclude that the domain specific applications are more specific for data mining. From above study it seems very difficult to design and develop a data mining system, which can work dynamically for any domain.

International Journal of Computer Science, Engineering and Information Technology (IJCSEIT), Vol.2, No.3, June 2012

## Authors

Presently Mr.Neelamadhab is working as a Asst Proff in I.T.Dept. He has done a post-graduate from Berhampur University, Berhamputr, India. He is a Life fellow member of Indian Society for Technical Education (ISTE). He is presently pursuing the doctoral degree In the field of Data Mining. He has total teaching experience of 9.5 years He has a total of 3 Research papers published in National / International Journals / Conferences into his credit. Presently he has also published 2 Books one is for Programming in C and other is Object Oriented Programming using C++. He has Received his MTech (computer 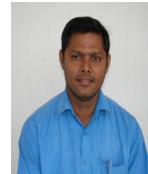 science) from Berhampur University Berhampur 2009,. His main research interests are Data warehousing and Mining, Distributed Database System. He is a resource scholar of CMJ university, Department of Computer Science, Meghalaya (Silong) ,India

Dr.Pragnayaban Mishra has completed his PhD in computer science from Sambalpur University, sambalpur (ODISSA) .He has published more than 8 journals of national and international .He is the life time member of ISTE.He is having 12 years of teaching experienced. He is presently guiding of 5 phd scholars. His research area is data ware housing and mining, Distributed System, Distributed data base and operating system, Neural network. Presently he is Professor and Head of Information and technology (IT) in GIET, Gunupur, Odisa. 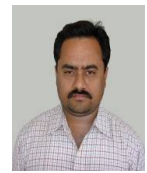

Mrs.Rasmita Panigrahi is currently working as a lecturer in the department of information and technology ,Gandhi Institute of Engineering and Technology .She received her MCA from BPUT,(Biju Patanaik University of Technology University ,Rourkela 2010 and she is pursuing MTech(Computer Science) in Berhampur University ,Berhampur Her main research interests are Data warehousing and Mining ,Distributed Database System, Designing and Algorithm. and cryptography .She has published two International/National paper and attended the several conferences . 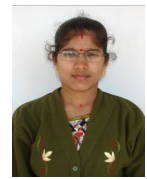